# FREQUENCY MULTIPLICATION IN TERAHERTZ BAND USING ALGAN/GAN PLASMONIC CRYSTALS

MICHAEL SHUR[*]

*Rensselaer Polytechnic Institute, Troy, New York 12180, USA)*
*Electronics of the Future, Inc., Vienna, VA 22181, USA*
*shurm@rpi.edu*

GREGORY AIZIN

*Kingsborough College, The City University of New York, Brooklyn, New York*
*gaizin@kbcc.cuny.edu*

*The plasma oscillations in high-mobility field-effect transistors (HEMTs) have emerged as a key physical mechanism for manipulating electromagnetic radiation in the sub-terahertz (sub-THz) and THz frequency ranges. These collective electron excitations can be excited and tuned electrically offering a compelling route to compact, integrable components for a wide range of next-generation technologies, including sixth-generation (6G) wireless networks, high-resolution biomedical and chemical spectroscopy, industrial process monitoring, and advanced security and defense systems. For these applications, plasmonic crystals—periodic arrays of many strongly coupled FET channels—are particularly promising. In this work, we report on a new class of collective excitations in plasmonic crystals, termed rotonic plasmons, which arise at plasmonic mode crossings and exhibit a parabolic dispersion law reminiscent of soft-mode and roton-like spectra. We show that uniform gate modulation across plasmonic crystal unit cells induces periodic variations in the sheet carrier concentration and, consequently, in the plasma frequency. This time-periodic modulation drives nonlinear plasmonic parametric resonances enabling RF-to-THz conversion. By solving the generalized Mathieu equation with damping, we demonstrate that high-amplitude gate pumping enables frequency multiplication and, at cryogenic temperatures (77 K), leads to parametric instabilities due to enhanced electron mobility. In plasmonic crystals with lower mobility, RF-to-THz conversion can instead be realized via periodic short-pulse excitation, a regime we introduce as Time-Domain Frequency Multiplication (TDFM). Investigations of AlGaN/GaN low–high plasmonic crystals confirm their potential as tunable, compact THz sources.*

*Keywords*: Plasmonic FETs, AlGaN/GaN heterostructures, THz frequency conversion, Grating-gate plasmonic crystals, 6G wireless sensing and imaging.

## 1. Introduction

Modern 5G smart phones operate at frequencies up to 40 GHz using GaAs Heterojunction Bipolar Transistors for power amplifiers. Higher power AlGaN/GaN technology is used for cell towers enabling 5G communication. However, rapidly expanding global network traffic will require the transition to higher frequencies and higher powers and AlGaN/GaN technology could be incorporated in the next generations of smart phones.

Plasma oscillations – i.e. oscillations of electron density in high-mobility field-effect transistors (HEMTs) have attracted intense research interest for their promise in sub-terahertz (sub-THz) and terahertz (THz) technologies. [1-3]

Plasmonic crystals—engineered arrays of hundreds or thousands of synchronized FET channels—can sustain resonant modes that enable THz phase detection and even compact THz radiation sources [4-8]. By combining many HEMT cells operating in synchrony they have potential of many orders of magnitude enhancement in oscillator power and detector sensitivity. Their potential applications span emerging 6G communications [1], biomedical [9] and chemical [10] sensing, industrial process control [11], and advanced defense systems. [12]

We recently demonstrated that a "low–high" grating-gate plasmonic FET architecture, with a high electron concentration in the ungated regions, can sustain strong resonant responses at room temperature across multiple material systems [13]. In these devices, the equilibrium electron density forms a periodic profile of gated and ungated regions with lengths $L_1$ and $L_2$ and electron densities $n_{01}$ and $n_{02}$, respectively. This paper focuses on AlGaN/GaN plasmonic structures with an arbitrary $n_{01}$ and $n_{02}$ (see Fig. 1).

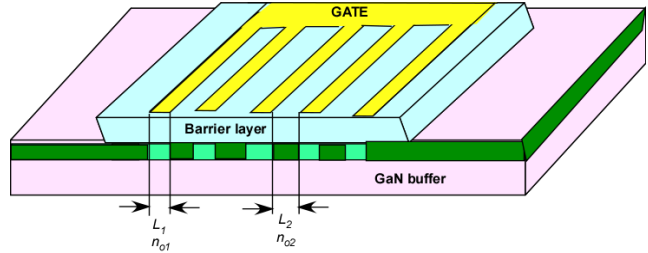

Fig. 1. One dimensional plasmonic crystal structure.

## 2. Roton plasmons and Mathieu equation

As shown experimentally [6] and theoretically [14] the mode crossover plays an important role in determining response of plasmonic crystals. Ref. [6] reported to the change from absorption to amplification at the cross point. At the crossover point, plasmons have parabolic dispersion law (the same as the roton dispersion law [15]) corresponding to the new type of plasmons that we call rotonic plasmons:

$$\omega_\pm = v_p M \Delta k \pm \sqrt{\delta^2 + \Delta k^2 v_p^2} . \qquad (1)$$

Here $v_p$ is the plasma velocity, $\Delta k = k - k_o$, $k$ is the wave vector, $k_o$ is the intersection point in $k$-space, $M = v_o / v_p$ is the Mach number, $v_o$ is the electron drift velocity, $v_p$ is the plasma velocity. and $\delta$ is the gap size. This spectrum reproduces the gap near the intersection point and the asymmetric shape of this gap at finite Mach numbers (see Fig. 2).

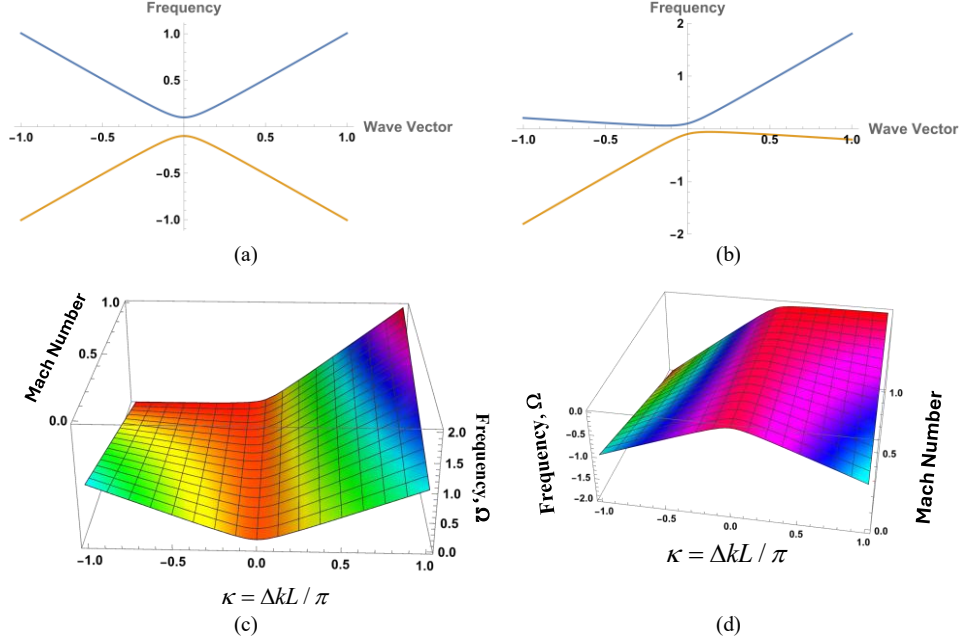

Fig. 2. Frequency $\Omega_\pm(\kappa, M) = \omega_\pm / \omega_p$, $\omega_p = 2\pi v_p / L_1$ is the fundamental plasma frequency, versus wave vector deviation from crossover point $\kappa = \Delta kL / \pi$ for $M=0$ (a) and $M=0.8$ (b); top mode (c) and bottom mode (d) versus $\kappa$ and $M$.

At $M=0$, this dispersion relationship is very similar to that in graphene, since in realistic graphene samples there's practically always a small energy gap on the order between 10 to 30 meV in vicinity of the Dirac point. It is also similar to an energy spectrum of narrow gap semiconductors where a large conduction band nonparabolicity makes the band dispersion to be very close to a nonideal Dirac cone. Fig. 2 shows the fitted plasmonic band spectrum near the bandgap, using Eq. (1). As seen in Fig. 2, a finite Mach number causes non-reciprocity.

At $M=0$, the plasmonic mode anticrossings result in parabolic spectra at vicinity of $k_o$, similar to roton spectrum and to a "soft-mode" spectrum near second order phase transition. [16, 17]. Therefore, the plasmonic oscillations at the crossing point could be described by the harmonic oscillator equation:

$$\frac{d^2 \delta N_\pm}{d\tau^2} + \gamma \frac{d\delta N_\pm}{d\tau} + \Omega_\pm^2 \delta N_\pm = 0. \qquad (2)$$

Here $\delta N_\pm = \delta n / n_1$, $\gamma = 1/(\omega_o \tau_m)$, where $\tau_m$ is momentum relaxation time, $\tau = \omega_o t$ is a normalized time, and $\Omega_\pm = \omega_\pm / \omega_o$, for the top and low branches in Fig. 2 are given by: [13]

$$\omega_+ = \omega_o \left[1 + \frac{L_2}{2L_1}\left(\sqrt{\frac{n_1^2}{n_2^2} + \frac{4L_1^2 M^2}{L_2^2}} - \frac{n_1}{n_2}\right)\right]. \qquad (3)$$

$$\omega_- = \omega_o \left[1 - \frac{L_2}{2L_1}\left(\sqrt{\frac{n_1^2}{n_2^2} + \frac{4L_1^2 M^2}{L_2^2}} + \frac{n_1}{n_2}\right)\right]. \qquad (4)$$

Here $\omega_o = (2\pi v_p)/L_1$, $v_p = \sqrt{q^2 n_{01}/m^* C}$, $q$ is the electronic charge, $m^*$ is effective electron mass, $C = K\varepsilon_o/d$ is the gate-to-channel capacitance per unit area, $\varepsilon_o$ is the dielectric permittivity of vacuum, $K$ is the dielectric constant of the barrier layer, $d$ is the effective gate-to-channel separation.

Temporal modulation of the parameters in the harmonic oscillator equation (Eq. 2) results in the parametric resonance at certain values of the modulation frequency. In contrast to a classical parametric resonance [18], we also consider a highly nonlinear parametric resonance, whose features are dramatically different from the prediction of classical theory. Below we demonstrate that such a nonlinear resonance leads to frequency multiplication and effective conversion of RF modulating signal into THz signal (see also Ref. 19)).

Here we only present the results for $M=0$. A more detailed analysis of the effect of the driving current ($M > 0$) will be presented elsewhere.

As mentioned above, we focus on a strongly nonlinear regime, when carrier density in the gated regions is modulated by a sinusoidal gate voltage at a lower frequency, $\delta\omega$: $\delta n = n_{01}(1 - A\cos\delta\omega t)$. In this case, the plasma frequencies at the crossing point are

$$\omega_{p\pm}^2 = \omega_\pm^2 B(\delta\omega t)$$

$$B(\delta\omega t) = \begin{cases} 1 - A\cos\delta\omega t & \text{if } A\cos\delta\omega t < 1 \\ 0 & \text{if } A\cos\delta\omega t > 1 \end{cases} \qquad (5)$$

Here $A$ is the modulation amplitude, and $B(\delta\omega t)$ is the frequency modulation function. Such a pumping allows all unit cells of a plasmonic crystal to operate in unison avoiding nonuniform excitation of plasmonic unit cells by driving drain-to-source current. Substituting $\Omega_{p\pm} = \omega_{p\pm}/\omega_o$ into Eq. (2) and assuming $\tau \to \infty$ leads to the Mathieu equation:

$$\frac{d^2 \delta N_\pm}{dT^2} + a\left(1 - \frac{2q}{a}\cos(2T)\right)\delta N_\pm = 0. \qquad (6)$$

Here $a = 4\omega_\pm^2/\delta\omega^2$, $2q/a = A$, $T = \delta\Omega\tau/2$.

The Mathieu equation is widely used for describing parametric resonances in mechanical and electrical systems [20-22]. In the next Section, we use the generalized Mathieu equation (accounting for damping) for the demonstration of frequency multiplication using "high-low" AlGaN/GaN plasmonic crystals. We show that a strong nonlinear parametric resonance supports frequency multiplication and Rf to THz frequency conversion. This device could compete with THz frequency multipliers relying on nonlinear properties of Schottky diodes [23].

### 3. Plasmonic response to a single excitation pulse under gate voltage modulation

Table 1 list materials and device parameters of AlGaN/GaN plasmonic crystals used in the simulation.

**Table 1.** Materials and geometry parameters

| Parameter | Unit | Value |
|---|---|---|
| Dielectric constant | | 10 |
| Effective mass | | 0.24 |
| Mobility 300 K, 77 K | m$^2$/Vs | 0.16, 2.5 |
| $L_1, L_2$ | m | $L_1, L_2 = 5\times10^{-8}$ and $1\times10^{-7}$ |
| Barrier layer thickness, $d$ | m | $2.5\times10^{-8}$ |
| $n_{10}, n_{20}$ | m$^{-2}$ | $10^{15}, 10^{16}$ |

Fig. 3 presents modulation function $B(\delta\omega t)$ and plasmonic oscillation waveforms after excitation pulse of $\delta n/n_{01}=0.1$ for excitation levels $A = 1$ and $A = 2$ for $\Omega_{p+}$ at 77 K.

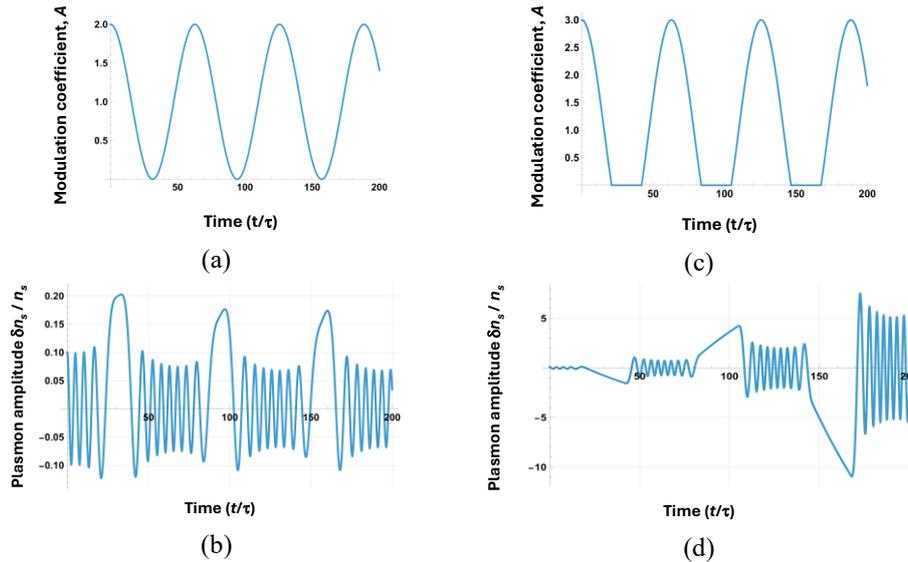

Fig.3. Modulation function and plasmonic oscillation waveforms after initial excitation pulse of $\delta n/n_{01}=0.1$ for $A = 1$ (a and b) and $A=2$ (c and d), for $\Omega_{p+}$ at 77 K ($L_1, L_2=5\times10^{-8}$ m). The unit of time is $1/\omega_o = 4.37\times10^{-14}$ s

As seen, at 77 K, a high enough excitation ($A = 2$) leads to an instability (with excitation level saturated by higher order nonlinearities, not considered here). In this regime, this device operates as a THz oscillator.

Fig. 4 presents the modulation function and plasmonic oscillation waveforms after initial excitation pulse of $\delta n/n_{01}=0.1$ for $M=0$ and excitation levels, $A = 1$ and $A=2$, for top mode $\Omega_{p+}$ for 300 K.

As seen, there's no instability at room temperature even for high pumping ($A = 2$).

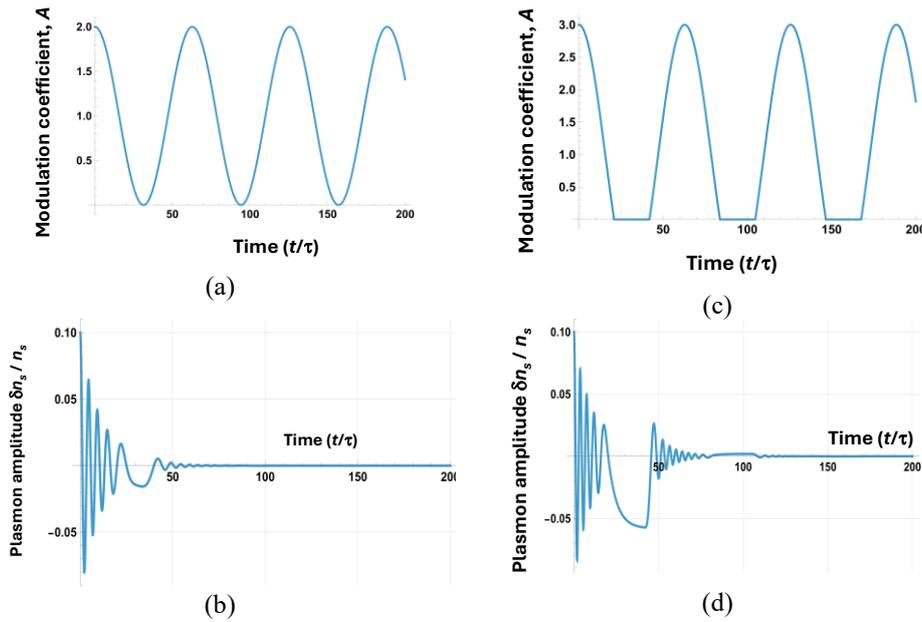

(a)          (c)

(b)          (d)

Fig. 4. Modulation function and plasmonic oscillation waveforms (amplitude versus time) after initial excitation pulse of $\delta n/n_{01}=0.1$ for $M=0$ and excitation levels, $A = 1$ (a and b) and $A=2$ (c and d), for top mode $\Omega_{p+}$ for 300 K ($L_1, L_2=5\times10^{-8}$ m). The unit of time in is $1/\omega_o = 4.37\times10^{-14}$ s.

## 4. Time Domain Modulated Source (TDMS)

The application of a short laser pulse that abruptly switches the conductivity of a material from very low to high by producing a large concentration of electron-hole pairs is classical method of generating THz radiation.[24, 25] This abrupt excitation and subsequent recombination of the generated electron-hole pairs results in an electromagnetic wave transient emitting an electromagnetic pulse with spectral components in the THz or sub-THz ranges. In this technique, low-temperature–grown gallium arsenide (LT-GaAs) is widely used as a THz emitter[26, 27] because it has an extremely short electron–hole recombination time. Similar effects have also been demonstrated in two-dimensional electron-gas heterostructures.[28, 29] Periodic excitation of THz emitter materials is widely used in THz spectroscopy (Time Domain Spectroscopy)- one of the key standard setups for THz characterization and imaging.[30, 31]

In this Section, we consider combining periodic photoconductive excitation with gate pumping—i.e., periodically modulating the gate voltage at a radio-frequency (RF) subharmonic, as illustrated in Figures 4a and 4c. In this new method, the period of the short optical pulses matches the period of the modulation function. We call this technique time-domain modulation, and we suggest it can serve as a novel THz source based on synchronized optical excitation and gate-voltage modulation.

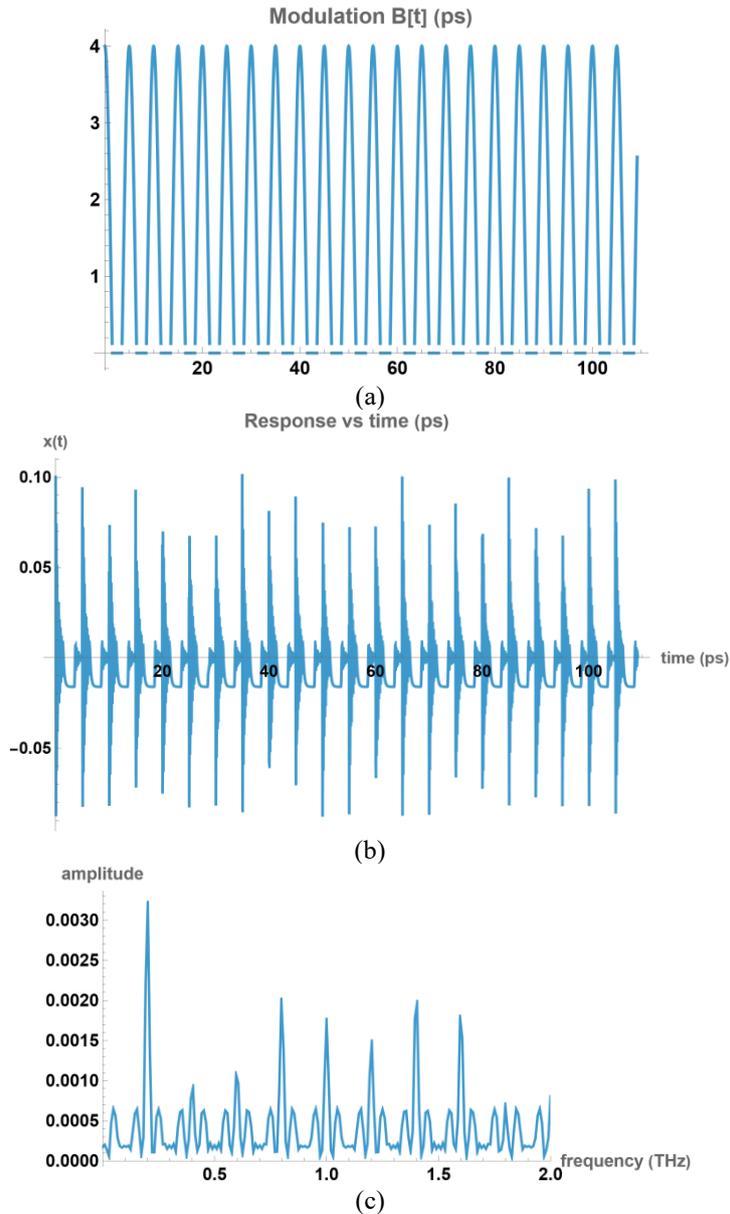

Figs. 5. Modulation function of plasmon frequency (a), plasmonic response (amplitude in units of $n_{10}$ ) (b) and the response's Fourier transform(c) for $L_1 = L_2 = 50$ nm and temperature 300 K. $A=3$.

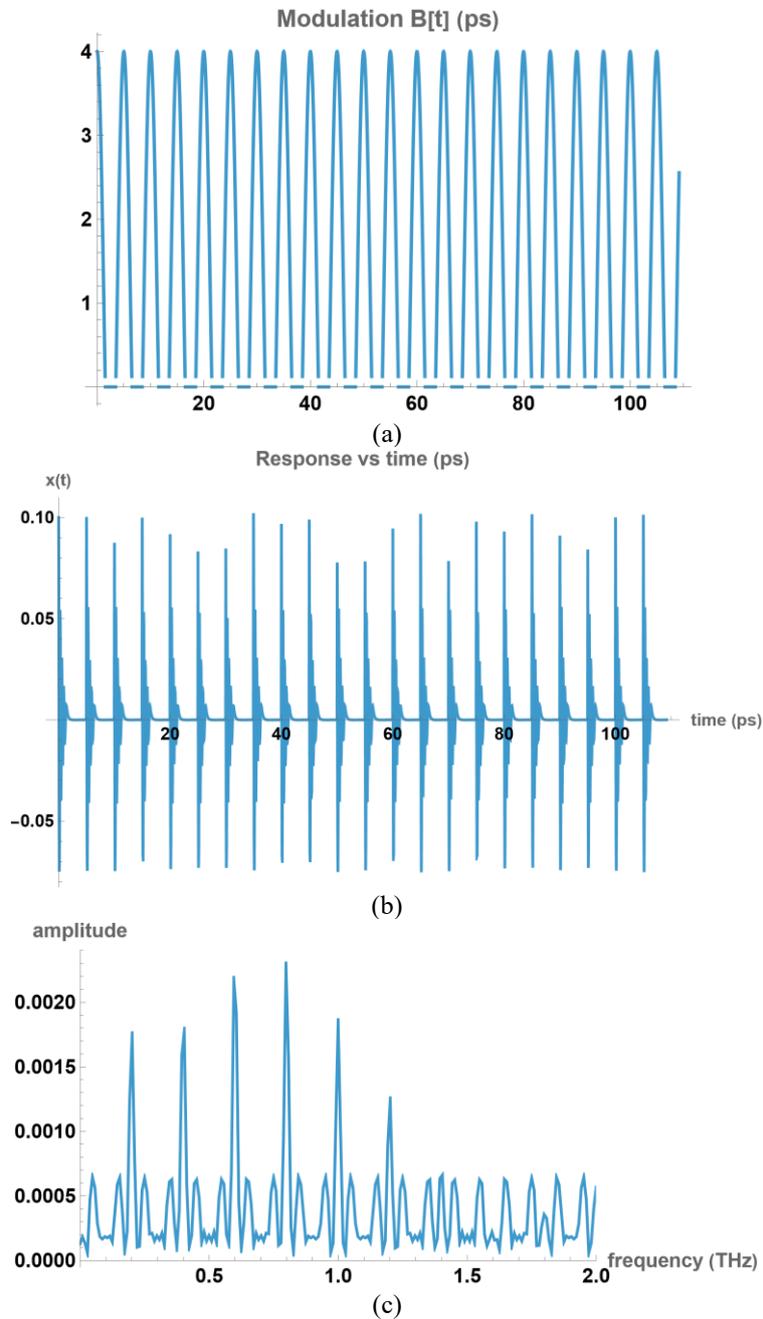

Figs. 6. Modulation function of plasmon frequency (a), plasmonic response (amplitude in units of $n_{10}$ ) (b) and the response's Fourier transform(c) for $L_1 = L_2 =100$ nm and temperature 300 K. A=3.

Figs. 5 and 6 show the plasmonic responses and response's Fourier transforms for $L_1 = L_2$ =50 nm and for $L_1 = L_2$ =100 nm, respectively. These waveforms are obtained for the regime when the excitation pulse is repeated with a period $T_B$ being equal to the period of the modulation function $B(2\pi t / T_B)$, i.e. pulse excitation and gate voltage modulation resonate. The value of $T_B$ is chosen to be on the order of $1/(\gamma \omega_o)$.

Fig. 7 compares the Fourier transforms for $L_1=L_2=50$ nm for $A=0$ and $A=3$.

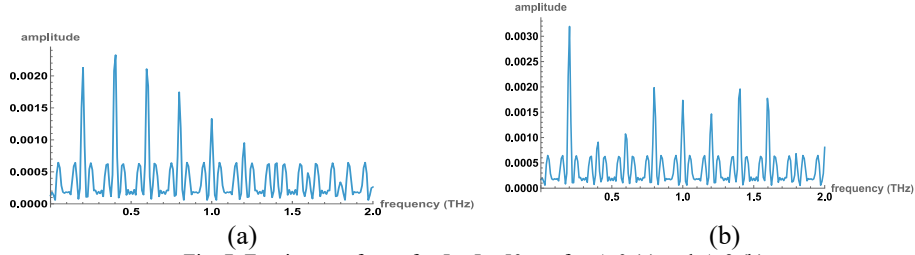

(a)                        (b)
Fig. 7. Fourier transforms for $L_1=L_2=50$ nm for $A=0$ (a) and $A=3$ (b).

As seen, the frequency of the plasmonic frequency modulation has a dramatic effect on the Fourier spectrum. In Fig. 7a (no plasma frequency modulation), the most prominent peaks are at 0.2 THz (which is the excitation frequency), 0.4 THz, and 0.6 THz. In Fig. 7b (strong plasma frequency modulation), the most prominent peaks are at 0.2 GHz and 0.8 THz. Hence, the device operates as an RF to THz converter.

The operating regime represented by Fig. 7b could be called Time Domain Frequency multiplication (compare with Transient Domain Spectroscopy). This regime could find multiple applications in compact, monolithically integrated THz components that can be incorporated into communication, sensing, and imaging systems.

## 5. Conclusions

In plasmonic crystals, the plasmonic dispersion at plasmonic mode crossings exhibits a parabolic dependence on the wave vector deviation from the crossover point. This parabolic dispersion law corresponds to the new type of plasmons that we call rotonic plasmons reminiscent of soft-mode and roton-like spectra. Modulating the gate voltage applied to all plasmonic crystal unit cells modulates the sheet carrier concentration, and, therefore, plasmonic frequencies. This time-periodic plasma frequency modulation supports plasmonic parametric resonances that exhibit new features, including RF to THz conversion in strongly nonlinear regimes. By solving the generalized Mathieu equation with damping, we demonstrated that modulation of plasma frequency by pumping the gate bias can result in frequency multiplication when the modulation amplitude is sufficiently high. At 77 K, such gate pumping leads to the onset of instability, since electron mobility is high at cryogenic temperatures. At 300 K, in plasmonic crystals with a relatively low mobility, RF to THz conversion could be realized by applying periodic short excitation pulses, in a regime similar to Time Domain Spectroscopy but with a strong plasma frequency modulation. We call this new regime of RF-THz conversion Time-Domain Frequency Multiplication (TDFM). We investigate this regime for AlGaN/GaN low-high plasmonic crystals and show that these crystals could serve as promising tunable sources of THz radiation. Their ability to operate at room temperature—and with enhanced performance at cryogenic temperatures—makes them attractive for integration into compact THz Time-Domain Frequency Multiplication (TDFM) systems enabling on-chip THz sensing and communication applications.